\documentclass[11pt]{article}
\usepackage{amsmath,amsfonts, amssymb,graphicx,geometry,authblk,color,soul,comment,hyperref} 
\begin{document}

\title{\textbf{Optical penetration depth and periodic motion of a photomechanical strip}}  
 
\author{Ameneh Maghsoodi$^{1}$\footnote{ Corresponding author: maghsoodi@usc.edu\\The research reported here was primarily carried out when AM was affiliated with Caltech.}}
\author{ Kaushik Bhattacharya$^{2}$}
\affil{$^{1}$ Department of Aerospace and Mechanical Engineering, University of Southern California, Los Angeles, CA 90089, USA}
\affil{$^{2}$Division of Engineering and Applied Science, California Institute of Technology, Pasadena, CA 91125, USA}
\maketitle

\begin{abstract}
\noindent {Liquid crystal elastomers (LCEs) containing light-sensitive molecules exhibit large reversible deformation when subjected to illumination. Here, we investigate the role of optical penetration depth on this photomechanical response. We present a model of the photomechanical behavior of photoactive LCE strips under illumination that goes beyond the common assumption of shallow penetration. This model reveals how the optical penetration depth and the consequent photomechanically induced deformation can depend on the concentration of photoactive molecules, their absorption cross-sections, and the intensity of illumination.  Through a series of examples, we show that the penetration depth can quantitatively and qualitatively affect the photomechanical response of a strip.  Shallow illumination leads to monotone curvature change while deep penetration can lead to non-monotone response with illumination duration.  Further, the flapping behavior (a cyclic wave-like motion) of doubly clamped and buckled strips that has been proposed for locomotion can reverse direction with sufficiently large penetration depth.  This opens the possibility of creating wireless light-driven photomechanical actuators and swimmers whose direction of motion can be controlled by light intensity and frequency.}\\

\end{abstract}

\maketitle

\begin{figure}
\begin{center}
  \includegraphics[width=6.4in]{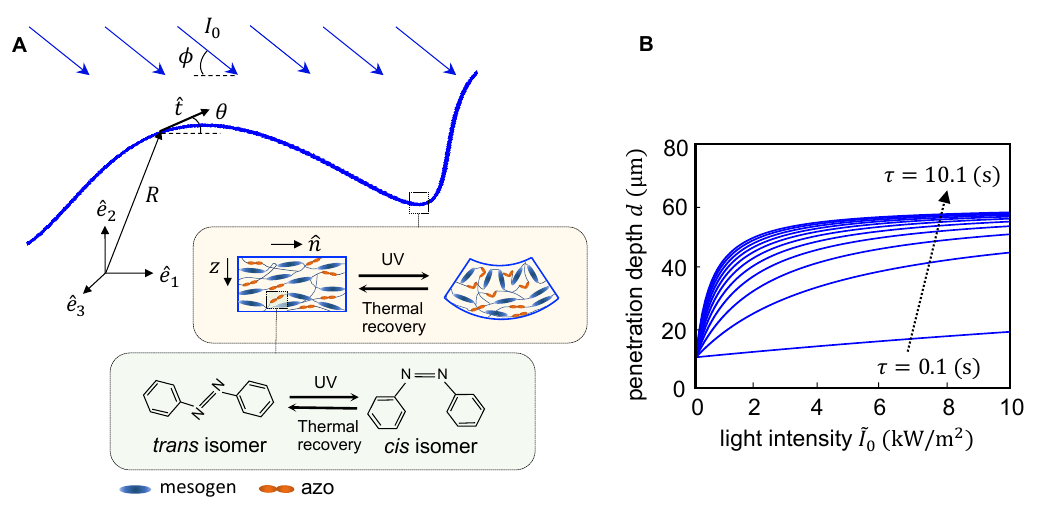}
  \caption{(A) Photomechanical strip under steady illumination and the underlying mechanism of deformation. (B) Optical penetration depth $d$ as a function of light intensity $\tilde I_0$ at different relaxation times $\tau$ =[0.1:1:10.1] (s).}
  \label{fig:fig1}
\end{center}
\end{figure}

\section{Introduction}
Liquid crystal elastomers (LCEs) containing light-sensitive molecules such as azobenzene exhibit fascinating reversible photomechanical behavior, in which, the light energy is converted to mechanical work \cite{white_2017}. Azobenzene molecules are mostly in their linear {\it trans} state in ambient conditions. Upon exposure to UV light, the molecule undergoes the isomerization process and transforms into a bent {\it cis} state. Thermal energy and exposure to visible light drive the reaction back to the {\it trans} state; see Fig.~\ref{fig:fig1}A. The photostationary state during illumination is a steady state between the forward {\it trans-cis} isomerization and the backward {\it cis-trans} recovery. 

In seminal work, Yu {\it et al.} \cite{yu_2003} use an isotropic genesis polydomain LCE film where domains of directors are randomly oriented.  When illuminated with plane polarized light, the isomerization of azobenzene reduces the orientational order of LC parallel to the direction of polarization, producing a spontaneous uniaxial contraction along the direction of polarization and a bending of the film.  Since then, there have been many advances, and much of the recent work has focused on either monodomain or patterned nematic LCEs \cite{gelebart2017making,white_2017}.  In a nematic LCE with nematic director $\mathbf{\hat n}$, the {\it trans-cis} isomerization of azobenzene reduces the orientational order of LC, producing a spontaneous uniaxial contraction along the direction of the nematic director. When a strip of LCE with a uniformly oriented director is exposed to light, absorption of light by isomerization decreases the intensity, and therefore the induced strain with thickness.  This in turn leads to a bending of the strip (Fig. \ref{fig:fig1}A). In this paper, we focus on nematic monodomain LCEs, but the considerations can easily be applied to the polydomain situation.

The amount of light-induced bending depends on the penetration depth, which in turn depends on material properties as well as the intensity and frequency of the light.  In particular, low light and high concentration lead to shallow penetration \cite{lall2023understanding,corbett2007linear}.  The overlap in the absorption spectra of the \textit{trans} and \textit{cis} states also leads to shallow penetration.  In this shallow penetration regime, the intensity of light decays exponentially with depth (Beer's law), and the photobending curvature increases monotonically with illumination time eventually reaching a steady state.  In contrast, in the deep penetration regime, the illumination profile deviates significantly from the exponential decay \cite{corbett2007linear,corbett2015deep}, and the light-induced curvature is no longer monotone with time \cite{corbett2015deep,van2008bending}.  Upon steady illumination, the curvature increases, reaches a peak, and then decreases to a non-zero steady state.  

Much of the modeling effort in the literature focuses on shallow penetration (e.g.,  \cite{warner2004photoinduced,corbett2007linear,korner2020nonlinear}).  Corbett, Xuan, and Warner \cite{corbett2015deep} developed a model of deep penetration and used it to study the free deformation of strips.  We build on their work in three ways.  First, their work did not consider the overlap of the absorption spectra of both the {\it trans} and {\it cis} states, and therefore limited absorption to the \textit{trans} state alone.  Second, we provide a very detailed analysis of the various factors that affect the penetration depth.  Third, we formulate the mechanics of strips with boundary conditions beyond the zero stress (free recovery formulation).  We compare various aspects of our model to available experimental observations.

We then study a doubly clamped buckled strip subjected to steady illumination (constant light intensity and direction).  Gelebart {\it et al} \cite{gelebart2017making} showed that such a strip undergoes a periodic flapping (wavelike) motion, and this motion was analyzed in detail in the regime of shallow penetration consistent with the experiment by Korner {\it et al.} \cite{korner2020nonlinear}.   Such motion can be used for locomotion \cite{gelebart2017making,maghsoodi2022light}.  We reproduce their results at shallow illumination, but find something remarkable in the deep penetration regime.  After an initial transient where the flapping motion is in the same direction, the motion reverses direction if the penetration depth is sufficiently large.  Since the penetration depth depends on the intensity of illumination as well as absorption cross-sections that depend on illumination frequency, this suggests the intriguing possibility for bidirectional motion.

\section{Optical penetration and deformation}

We consider a long strip or beam with a rectangular cross-section subjected to the incident light intensity $ I_0$, as shown in Figure \ref{fig:fig1}A. The nematic director is parallel to the long axis of the strip, known as homogeneous strip. The illumination direction and the motion are assumed to be in a plane along one of the main axes of the cross-section.  Thus, we can treat this in two dimensions.  We introduce $S$ to be the coordinate along the centerline of the strip in the reference configuration and $z$ to be the coordinate along the thickness of the strip.  Since the strip is very long, we first fix $S$ to study the penetration and the spontaneous deformation at coordinate $S$.  We then consider the entire strip.

\subsection{Optical penetration}
We fix $S$ and suppress it from the notation.  Let $n_c$ denote the fraction $N_c/N$ of azobenzene molecules in the {\it cis}  state where $N_c$ is the number density of azobenzene molecules in the {\it cis} state and $N$ is the total number density of azobenzene molecules. Let $n_t$ and $N_t$ analogously denote the fraction and number density of azobenzene molecules in the {\it trans} state.  The fraction of {\it cis} molecules changes due to isomerization induced by light and due to thermal relaxation. So, at a depth $z$ at time $t$, the rate of change of the {\it cis} volume fraction is
\begin{eqnarray}
\frac{\partial n_c(z,t)}{\partial t} =  \frac{a_t}{E_p} I(z,t) n_t(z,t)-   \frac{a_c}{E_p}  I(z,t) n_c(z,t) - \frac{1}{\tau} (n_c(z,t) - n_{c_0}), \label{eq:conc}
\end{eqnarray}
where $I$ is the normal light intensity (units of energy per area perpendicular to $z$ per time) at depth $z$ at time $t$, $a_t$ and $a_c$ are the absorption cross-section (units of area per molecule) of the {\it trans} and {\it cis} states, respectively, $E_p$ is the energy per photon, $\tau$ is thermal relaxation time, and $n_{c_0}$ is the equilibrium concentration of {\it cis} molecules in the absence of any illumination.  The first term describes the rate of isomerization of the {\it trans} molecules, the second term is the rate of isomerization of the {\it cis} molecules, and the final term is the thermal relaxation.

\begin{table}
\caption{Typical values used for the demonstration of the model.}
\centering
\begin{tabular}{l l l }
\hline
Parameter & Value & Description\\
\hline
$E_p$ & $5\times 10^{-19}$ J  \cite{guo2022regimes} & energy per photon at 365nm\\ 
$N$ & $10^{26}$ m$^{-3}$  \cite{guo2022regimes} & number density of azo molecules\\ 
$a_t$ & $10^{-21}$ m$^2$ \cite{guo2022regimes} & absorption cross-section of \textit{trans} isomer\\
$a_c$ & $10^{-22}$ m$^2$ &  absorption cross-section of \textit{cis} isomer\\
$n_{c_0}$ & $10^{-3}$ & equilibrium concentration of \textit{cis} molecules\\
$\rho$ & 1.4  g.cm$^{-3}$ \cite{smith2014designing}  & mass density of elastomer\\
$w$ &  1 mm \cite{smith2014designing}  & strip width\\
$L_0$ &  15 mm \cite{smith2014designing}  & strip length\\
 $H$ & $15 \mathrm{\mu m}$ to $1000 \mathrm{\mu m}$ & strip thickness\\
 $Y$ & $4$ GPa \cite{smith2014designing}& Young's modulus \\
 $\lambda$ & $0.05$ \cite{corbett2015deep}& \textit{cis}-strain proportionality\\
 $\tau$ & 0.2s to 10s & \textit{cis} relaxation time\\
 $\phi$ & $0^\circ$ to $90^\circ$  & illumination dangle\\

\hline
\end{tabular}
\label{tab:param}
\end{table}

As light is absorbed by the azobenzene molecules, its intensity diminishes with depth, 
\begin{eqnarray}
\frac{\partial I(z,t)}{\partial z} =  - a_t N I(z,t) n_t(z,t) -   a_c N  I(z,t) n_c(z,t). 
\end{eqnarray}
It is useful to introduce a penetration depth and rewrite this equation in terms of this penetration depth.  There are many conventions that are used to define the penetration depth.  We do so in the photo-stationary state when the distribution of molecules $\bar n_c(z)$ and $\bar n_t(z)$ are independent of time.  We rewrite (\ref{eq:inten}) as
\begin{eqnarray}
\frac{\partial I(z,t)}{\partial z}= \frac{-I(z,t)}{d} \frac{a_t n_t(z,t) + a_c n_c(z,t)}{a_t \bar {n}_t(0) + a_c \bar{n}_c(0)},  \label{eq:inten}
\end{eqnarray}
where we set the penetration depth $d= 1/ (N a_t \bar {n}_t(0) + N a_c \bar{n}_c(0))$.  Now, we can use (\ref{eq:conc}) in the photo-stationary state when the left-hand side is zero and the condition $n_c + n_ t =1$ for all $z$ and $t$ to compute $\bar n_c(z)$ and $\bar n_t(z)$.  Setting $z=0$, we conclude that the penetration depth
\begin{equation}
d=\frac{ (a_t + a_c) \tau \tilde I_0 + E_p}{N \left(2a_t a_c \tau \tilde I_0 + E_p a_t  + E_p n_{c_0}(a_c - a_t)\right)}, \label{eq:d}
\end{equation}
where $\tilde I_0$ is the normal intensity of light incident at the surface of the strip ($z=0$). Equations (\ref{eq:conc},\ref{eq:inten}) with $d$ defined in (\ref{eq:d}) describe the concentration evolution and the penetration of light.  These reduce to the equations of Corbett, Xuan, and Warner \cite{corbett2015deep} if we take $a_c=0$.

Figure \ref{fig:fig1}B shows penetration depth as a function of normal light intensity $\tilde I_0$ and relaxation time $\tau$ while other parameters remain constant and as given in Table \ref{tab:param}.  We see that the penetration depth increases with both the intensity of light and relaxation time.  We can understand the dependence on illumination by looking at the asymptotic limits.   In the limit of low illumination, 
\begin{equation} \label{eq:shallow}
d=\frac{ 1}{N \left(a_t + n_{c_0}(a_c - a_t)\right)} \approx \frac{ 1}{N a_t},
\end{equation}
The approximation follows since it is typical that the equilibrium {\it cis} concentration $n_{c_0}$ is small at ambient temperature and the absorption cross-sections overlap resulting in $a_c$ being a significant fraction of $a_t$.  So, the penetration depth is small especially if the azobenzene concentration is high.  Further, $n_c \approx 0$ and $n_t \approx 1$,  resulting in the well-studied Beer's shallow penetration limit \cite{warner2004photoinduced} where the penetration depth $d$ is small and the illumination decays exponentially.  Conversely, in the limit of high illumination (bleached limit),
\begin{equation} \label{eq:bleach}
d=\frac{ 1}{2N  } \left(\frac{1}{a_c} + \frac{1}{a_t} \right),
\end{equation}
Comparing (\ref{eq:shallow}) and (\ref{eq:bleach}), and recalling that the absorption cross-section of {\it cis} is smaller than the absorption cross-section of {\it trans} molecules ($a_c < a_t$), we find that the depth of penetration increases with illumination.  Further one can get deeper penetration as the absorption spectra are separated.

Finally, we comment on the role of the frequency of the light.  This affects the penetration depth through the photon energy as well as the absorption cross-sections that are frequency-dependent.

\subsection{Light-induced stretch and curvature}

We now examine how the penetration of light affects the spontaneous stretch and curvature of the strip.  We fix $S$ and suppress it in the notation.  It is convenient to change variables $Z=H/2-z$ where $Z$ denotes the distance from the centerline of the cross-section. The photo-isomerization of the azobenzene molecules leads to a spontaneous strain (strain when the stress is zero) in the material.  This strain $\varepsilon_0$ typically exhibits a linear relationship with the concentration of the {\it cis} molecules (e.g., \cite{van2007glassy}), $\varepsilon_0 (Z,t) = - \lambda n_c(Z,t)$ where $\lambda$ is a material constant depending on the alignment of the nematic directors in the nematic LCE. We use the negative sign since photo-isomerization leads to a contraction, and this allows $\lambda >0$.  The axial stress may now be written as the 
\begin{equation}
\sigma(Z,t) = Y \left(\varepsilon(Z,t) - \varepsilon_0 (Z,t) \right)
\end{equation}
where $Y$ is the Young's modulus and $\varepsilon$ is the total strain at depth $Z$ at time $t$.  The total axial force $T$ and total moment $q$ about the center point across the cross-section may now be calculated to be
\begin{align}
T(t)&=\int_{-H/2}^{H/2}  Y(\varepsilon(Z,t)-\varepsilon_0(Z,t)) \mathrm{d}Z,  \label{eq:f1} \\
q(t)&=\int_{-H/2}^{H/2}  Y(\varepsilon(Z,t)-\varepsilon_0(Z,t)) Z \mathrm{d}Z. \label{eq:q1}
\end{align}
The strain $\varepsilon$ may be related to the overall shape of the beam using the classical assumption that plane cross-sections remain plane in a beam.  Let $\epsilon(t)$ denote the axial strain and $\kappa(t)$ denote the curvature associated with the centerline at the point $S$ under consideration.  Then, 
\begin{equation}
\varepsilon(Z,t) = \epsilon(t) + Z \kappa(t)
\end{equation}
Substituting this back, in (\ref{eq:f1},\ref{eq:q1}), we conclude
\begin{align}
T(t)&=  YH (\epsilon(t) - \epsilon_0(t)), \label{eq:T} \\
q(t)&= \frac{YH^3}{12} (\kappa(t) - \kappa_0(t)) \label{eq:q}
\end{align}
where the spontaneous stretch and curvature of the cross-section are
\begin{align}
\epsilon_0(t) &= - \frac{\lambda}{H} \int_{-H/2}^{H/2}  n_c(Z,t)\mathrm{d}Z, \label{eq:ep0}\\
\kappa_0(t)&=\kappa_{in} - \frac{12\lambda}{H^3}\int_{-H/2}^{H/2} n_c(Z,t) Z \mathrm{d}Z \label{eq:k0}.
\end{align}
where $\kappa_{in}$ is the intrinsic curvature of the beam in the reference configuration.  

\begin{figure}
\begin{center}
  \includegraphics[width=6.5in]{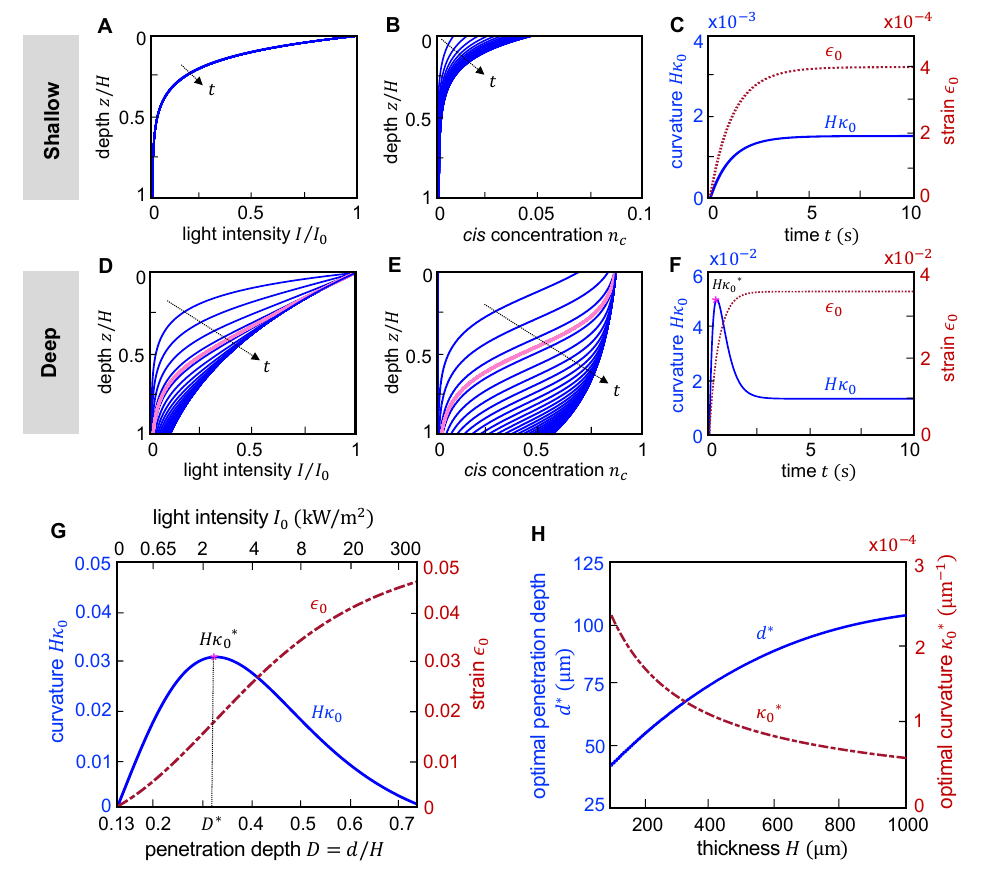}
  \caption{Light-induced spontaneous stretch and curvature.  (ABC) The case of shallow penetration ($I_0=50$W/m$^2$ and $D=0.14$): (A) Evolution of light intensity with depth, (B) Evolution of {\it cis} concentration with depth, (C) Evolution of spontaneous stretch and curvature with time.  (DEF) The case of deep penetration ($I_0=15$kW/m$^2$ and $D=0.58$): (D) Evolution of light intensity with depth, (E) Evolution of {\it cis} concentration with depth, (F) Evolution of spontaneous stretch and curvature with time.  The pink curve in both (D) and (E) corresponds to the peak spontaneous curvature.  (G) The light-induced spontaneous strain $\epsilon_0$ and non-dimensional curvature $H \kappa_0$ at different illumination intensities, and thereby, different penetration depths $D=d/H$. In (A-G), $H=150 \mu$m. (H) The optimal penetration depth $d^*$ and corresponding optimal curvature at different thicknesses $H$. In (A-H), $\tau=1$s.} 
\label{fig:epk}
\end{center}
\end{figure}

Figure \ref{fig:epk} shows details of the evolution of the spontaneous stretch and curvature, and how they depend on various parameters (other parameters as given in Table \ref{tab:param}).  We consider a strip in thermal equilibrium and subject it to light with normal intensity $\tilde I_0$.  We solve (\ref{eq:conc}, \ref{eq:inten}) simultaneously to find the evolution of the {\it cis} concentration $n_c(Z,t)$.  We then use it in (\ref{eq:ep0}, \ref{eq:k0}) to find the spontaneous stretch $\epsilon_0$ and curvature $\kappa_0$.  

Figures  \ref{fig:epk}A-C show the case of shallow penetration when the incident illumination $I_0=50$W/m$^2$ ($D=0.14$).  The intensity of light decays exponentially with depth and changes very little with time (Figure \ref{fig:epk}A).  The {\it cis} concentration also decreases exponentially with depth, but the concentration at the surface increases monotonically with time till it reaches a steady photo-stationary value (Figure \ref{fig:epk}B).  Consequently, both the spontaneous strain and curvature increase with time till they reach a steady photo-stationary value (Figure \ref{fig:epk}C).  Further, the photo-stationary strain is extremely small (100 times smaller than the point-wise maximum photo-strain).  All of this is consistent with the traditional models \cite{lall2023understanding,warner2004photoinduced,corbett2007linear,korner2020nonlinear}, and observations \cite{yu_2003,white_2017,gelebart2017making}.   

The case of deep penetration is quite different as shown in Figures \ref{fig:epk}D-F when the incident illumination $I_0=15$kW/m$^2$ ($D=0.58$).  We see that the light intensity decays exponentially when first illuminated, but gradually penetrates deeper till it reaches a photo-stationary profile where light is transmitted through the other surface (Figure \ref{fig:epk}D).  The early profiles of {\it cis} concentration also decay exponentially, but the surface eventually saturates and the concentration becomes more uniform through the thickness (Figure \ref{fig:epk}E).  Figure \ref{fig:epk}F shows the evolution of the overall spontaneous strain and curvature.  We see that the spontaneous strain increases monotonically reaching a photo-stationary value that is very high (80\% of the point-wise maximum photo-strain).  The spontaneous curvature initially rises, but then reaches a peak and gradually diminishing to a photo-stationary value.  

Figure \ref{fig:epk}G shows how the spontaneous stretch and curvature at the photo-stationary state (at long time) depends on the intensity of illumination, equivalently penetration depth.  We see that the photo-stationary spontaneous stretch increases monotonically, but the photo-stationary spontaneous curvature has a non-monotone behavior.  Figure \ref{fig:epk}H shows the critical value of the penetration depth at which the peak photo-stationary spontaneous curvature is attained an the corresponding value of the maximum curvature as a function of the thickness of the strip or beam.

\subsection{Photomechanical model of a strip}

We are now in a position to relate the effect of illumination studied in the previous sections to the overall shape of the strip or beam. Let $\mathbf R(S,t)$ denote the current position of the point $S$ on the centerline of the beam at the time $t$ -- see Figure \ref{fig:fig1}.  The tangent to the centerline is $\hat{\mathbf{t}} = | \frac{\partial \mathbf{R}}{\partial S}|^{-1} (\frac{\partial \mathbf{R}}{\partial S})$ and the axial strain $\epsilon(S,t) =  | \frac{\partial \mathbf{R}}{\partial S}| - 1$.  Let $\theta $ denote the angle that $\hat{\mathbf{t}}$ makes to the horizontal: then the curvature $\kappa(S,t) = \frac{\partial \theta}{\partial S} (S,t)$.  Finally, we assume that the beam in not shearable: this means that cross-sections are constrained to remain normal to $\hat{\mathbf t}$.  So $\hat{\mathbf t}$ also describes the normal to the cross-section.

The balance of linear and angular momenta may be written as 
\begin{align} 
m \frac{\partial^2 \mathbf R(S,t)}{\partial t^2} & = \frac{\partial \mathbf f(S,t)}{\partial S} +\mathbf f_\text{ext}(S,t)  
\label{eq:lin}\\
\Pi \frac{\partial^2 \theta(S,t)}{\partial t^2} &= 
\frac{\partial q(S,t)}{\partial S}+  (1+\epsilon(S,t)) (\hat {\mathbf t}(S,t) \times \mathbf f(S,t)). \hat{\mathbf e}_3 + q_\text{ext}(S,t)
\label{eq:ang}
\end{align}
where $\mathbf{f}$ and $q$ are the internal force and moment respectively, $\mathbf{f}_\text{ext}$ and $q_\text{ext}$ are the external force and moment per
unit length applied to the system respectively, $m$ and $\Pi$ the mass and rotational moment of inertia per unit length respectively.  The internal forces and moments are related to the deformation.    It is convenient to write the internal force  in terms of its normal and tangential (axial) components,  $\mathbf{f} = N\hat{\mathbf n}+ T \hat{\mathbf t}$.  The constraint that the beam is not shearable means that the shear force or normal component $N$ is indeterminate.  The tangential or axial force depends on the strain and spontaneous light-induced strain as described by (\ref{eq:T}) while the internal moment depends on the curvature and spontaneous curvature as described by (\ref{eq:q}).

Finally, the beam is illuminated with a light of intensity $I_0$ in the direction $\phi$ relative to the horizontal axis.  Then, at any point $S$ at time $t$, the normal intensity of illumination at the surface 
\begin{equation} \label{eq:ill}
\tilde I_0 = I_0 \cos \left(\pi/2 - \phi - \theta \right).
\end{equation}
This relates the illumination incident on any section to the shape of the beam.
\vspace{\baselineskip}

In summary, at every time step, given the current shape of the strip, we update the distribution of {\it cis} molecules and penetration of light at each $S$ by integrating (\ref{eq:conc},\ref{eq:inten}) subject to the normal incident intensity $\tilde I_0$, use it to compute the axial force and moment at each $S$ according to (\ref{eq:T}) and (\ref{eq:q}) respectively, and then integrate (\ref{eq:lin}, \ref{eq:ang}) subject to boundary conditions and  the non-shearing constraint to find the new shape of the beam.  We use an implicit discretization in time and finite difference in depth to update (\ref{eq:conc},\ref{eq:inten}) and a numerical method discussed in \cite{gatti2002physical,maghsoodi2016first,gobat2002generalized} to update (\ref{eq:lin},\ref{eq:ang}).

\section{Examples}

We study two examples.  The parameters are shown in Table \ref{tab:param} unless otherwise specified.

\subsection{Photobending of a clamped-free strip} 
\begin{figure}[t]
\centering
\includegraphics[width=5in]{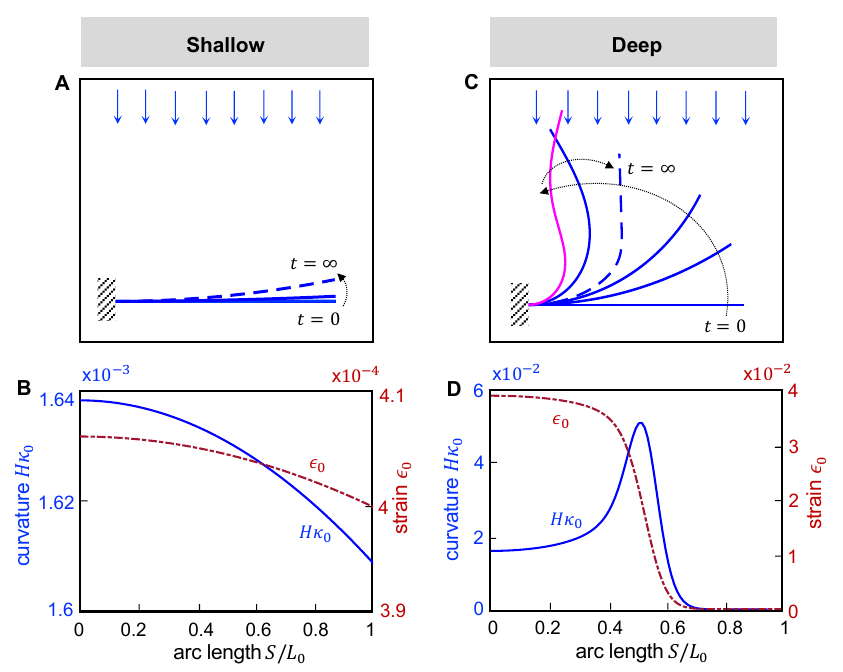}
   \caption{Dynamics of the photomechanical response of a clamped-free strip with homogeneous molecular alignment. (AB) Shallow penetration ($I_0=50$W/m$^2$ and $D=0.14$): (A) Snapshots of the strip at various times. Photo-stationary state is shown by dashed line. (B) Distribution of spontaneous strain and curvature along the length at the photo-stationary state in (A).     (CD) Deep penetration ($I_0=15$kW/m$^2$ and $D=0.58$): (C) Snapshots of the strip at various times. The pink curve corresponds to the maximum bending, and the dashed line indicates photo-stationary state. (D) Distribution of spontaneous strain and curvature along the length at the photo-stationary state in (C). In these simulations, $\phi=90^\circ$, $H=150 \mathrm{\mu m}$, and $\tau=1$s.} 
\label{fig:CF}
\end{figure}

Our first example considers a homogeneous strip illuminated from above while one end is clamped and the other is free.   Since one end is free, there is no net force or moment acting at any cross-section.   Thus, this is an example of free deformation as studied experimentally \cite{van2008bending,wang2011photomechanical,lee2012photomechanical}, and analyzed by Corbett {\it et al.} \cite{corbett2015deep}.  

The results are shown in Figure \ref{fig:CF}. The case of shallow penetration  ($I_0=50$W/m$^2$ and $D=0.14$) is shown in Figures \ref{fig:CF}(AB).  The free end of the strip bends up toward the light source monotonically till reaches its final photo-stationary shape (Fig. \ref{fig:CF}A and Movie 1 in Supplement). The spontaneous curvature and strain decrease monotonically from the clamped to the free end as also observed in \cite{wang2011photomechanical}: this is to be expected since the deformation of the strip changes the incident angle of the light.  The case of deep penetration ($I_0=15$kW/m$^2$ and $D=0.58$) is shown in Figures \ref{fig:CF}(CD).  The free end of the strip again bends up towards the light, but much more than in the case of shallow penetration.  However, the motion is not monotone: it reaches its peak as shown by pink curve in Figure \ref{fig:CF})C, but then relaxes back to the photo-stationary state as shown by dashed line (see Movie 1 in Supplement).  In the photo-stationary state, much of the strip is vertical (so that no light is incident normal to the strip) and the deformation is confined to the clamped end where the incident light is normal to the strip.  This is consistent with the spontaneous strain and curvature shown in Figure \ref{fig:CF}D.  These results are consistent with the analysis of Corbett {\it et al.} \cite{corbett2015deep}, and the observations of Oosten et al. \cite{van2008bending} who observed that the films with a low concentration of azobenzene (deep penetration) reach a maximum in bending and then slowly relax back to a bent state over time while exposed to illumination .

\subsection{Periodic motion of a doubly clamped strip} 
\begin{figure*}
\begin{center}
  \includegraphics[scale=0.85]{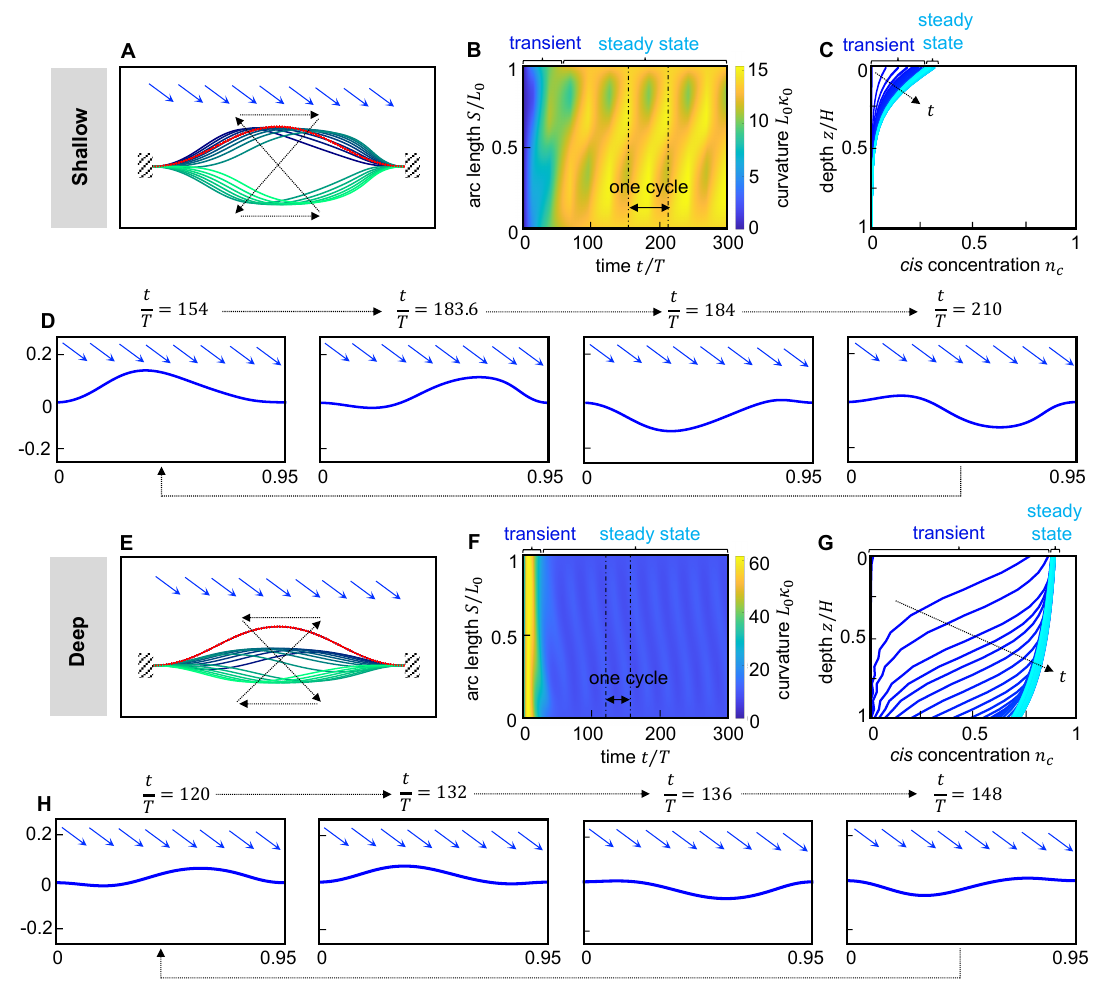}
  \caption{Dynamics of the photomechanical response of a doubly-clamped buckled strip with homogeneous molecular alignment. (A-D) correspond to shallow penetration ($I_0=630$W/m$^2$ and $D=0.2$), and (E-H) correspond to deep penetration ($I_0=31$kW/m$^2$ and $D=0.65$). (A) and (E) present the flapping motion per cycle; frequency $f=3.2$Hz for the shallow penetration, and frequency $f=6.1$Hz for the deep penetration. The red curves indicate the initial conformation of the strip when the light is off. (B) and (F) present the time evolution of non-dimensional spontaneous curvature $L_0 \kappa_0$ along the strip arc length. (C) and (G)  present the time evolution of $cis$ concentration $n_c$ across the thickness. (D) and (H) present the snapshots of light-induced deformations of the strip during one cycle. In these simulations, $T=\sqrt {\frac{12 \rho L_0^4}{Y H^2}}$ is proportional to the natural period of the strip. Also, $\phi=45^\circ$, $\tau=0.2$s, $L=0.95L_0$, and $H=15 \mathrm{\mu m}$.}   
\label{fig:CC}
\end{center}
\end{figure*}

The second example we study is a pre-stressed homogeneous strip under steady illumination. Gelebart {\it et al.} \cite{gelebart2017making} showed that a doubly clamped buckled strip undergoes a periodic motion under steady illumination.  The penetration depth in these experiments was shallow, and Korner {\it et al.}  \cite{korner2020nonlinear} studied studied this and other snap-through buckling phenomena using a model in the Beer's limit.  In this work, we examine if and how the phenomenon would differ in deep penetration.

We consider a strip of length $L_0$ that is straight in the absence of any illumination or stress (so $\kappa_{in}=0$). We clamp both ends at a distance $L<L_0$ from each other. In doing so, the strip buckles either upward or downward since there are two fundamental buckling modes. We consider one of the two states, say, the strip buckled up. We illuminate the strip with a uniform light source at a constant angle $\phi$ and compute the evolution.  The results are shown in Figure  \ref{fig:CC} (as before, simulation parameters are listed in Table \ref{tab:param})

Figure  \ref{fig:CC}A-D show the results of the case of shallow penetration ($I_0=630$W/m$^2$ and $D=0.2$).  These are consistent with those of Korner {\it et al.}  \cite{korner2020nonlinear} and observations of Gelebart {\it et al.} \cite{gelebart2017making}.   Figure \ref{fig:CC}A provides a number of snap-shots of the motion (see Movie 2 in Supplement), and Figure \ref{fig:CC}B shows the evolution of the non-dimensional spontaneous curvature $L_0 \kappa_0$ along the length of the strip. After a very short initial transient, the illuminated strip settles into a periodic `flapping motion'. Figure \ref{fig:CC}D shows four snap-shots of the steady state:  the up bump moves slowly from left to right (away from the light source), then jumps suddenly  from up-right to the down-left, then the down bump moves slowly from left to right, and finally jumps back to the up-left position.  Figure \ref{fig:CC}C shows the evolution of the concentration $n_c$ at the middle of the strip.  It decays exponentially from the illumination surface, with a periodic amplitude at long times.

The situation with deep penetration ($I_0=31$kW/m$^2$ and $D=0.65$) is quite different as shown in Figure  \ref{fig:CC}E-H. Figure \ref{fig:CC}E provides a number of snap-shots of the motion (see Movie 2 in Supplement), and Figure \ref{fig:CC}F shows the evolution of the non-dimensional spontaneous curvature $L_0 \kappa_0$ along the length of the strip.  After a very short initial transient, the beam seems to move to a periodic `flapping motion' as before.  However, there is another transient before which the beams settles in a periodic `flapping motion': {\it however, the direction of the periodic flapping motion in this deep penetration situation is opposite to the direction of the periodic flapping motion in the previous shallow penetration situation}.  This is illustrated in the four snap-shots in Figure \ref{fig:CC}H. The up bump moves slowly from right to left (towards the light source), then jumps suddenly  from up-left to the down-right, then the down bump moves slowly from right to left, and finally jumps back to the up-right position.  The reason for this opposite motion is clear by looking at the concentration $n_c$ shown in Figure \ref{fig:CC}G.  Due to the deep penetration of light, the concentration has an inverse exponential profile and oscillates about this profile.  One final observation; note in Figure \ref{fig:CC}E that the amplitude of the buckle is smaller after illumination than in the initial state.  This is because one has significant contractive spontaneous strain in this deep penetration case.

\begin{figure}
\begin{center}
  \includegraphics[scale=1]{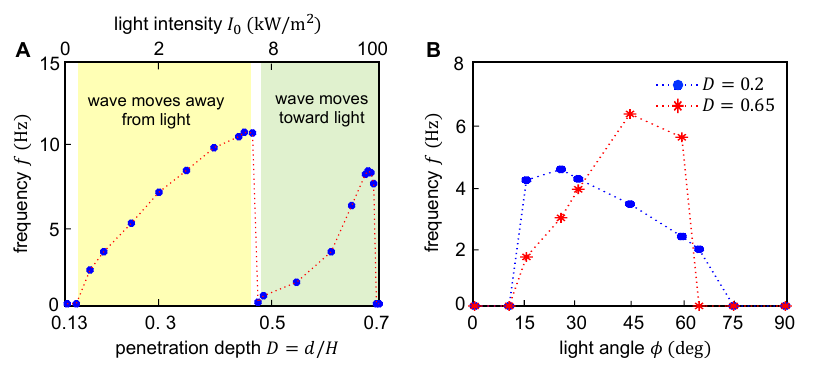}
\caption{Steady state frequency of flapping motion of strip: (A) at various penetration depth $D=[0.13,0.70]$ corresponding to the light intensity $I_0=[0,100]$ kW/m$^2$. The illumination angle is $\phi=45^\circ$. (B) at various illumination angles $\phi=[0^\circ,90^\circ]$ for penetration depth $D=0.2$ ($I_0=630 \mathrm{W/m^2}$) and $D=0.65$ ($I_0=31 \mathrm {kW/m^2}$). In these simulations, $\tau=0.2$s, $L=0.95L_0$, and $H=15 \mathrm{\mu m}$.}  
 
\label{fig:CC_freq}
\end{center}
\end{figure}

We now examine in greater detail the change of the flapping motion with penetration depth (illumination intensity).  Figure \ref{fig:CC_freq}A shows the steady state frequency of flapping motion of the strip at various penetration depths and $\phi=45^\circ$.   The photomechanical response exhibits two distinct regimes, depending on the direction of steady-state flapping motion; regime I, in which the wave moves away from the light source (see, for instance, Fig. \ref{fig:CC}A-D), and regime II in which the wave initially moves away from light and then reverses its direction and eventually moves toward light source at steady state (see, for instance, Fig. \ref{fig:CC}E-H). In this example, $D^c=0.48$ corresponding to $I^c_0=7 \mathrm{kW/m^2}$ is the critical penetration depth separating two regimes. At the critical $D^c$, the rate of $cis$ concentration $\dot{n_c}$ at the upper and lower layers becomes balanced after an initial transient, and consequently, the strip eventually stops at a photostationary state. Figure \ref{fig:CC_freq}A also reveals that, at each regime, the frequency increases gradually with $D$ to reach a maximum value and then dramatically decreases to zero. Importantly, each regime requires a critical light intensity to initiate a periodic photomechanical motion. 

We repeat calculations for various illumination angles at each regime. The results are summarized in Fig. \ref{fig:CC_freq}B. In each regime, there exists a specific angle where the frequency of flapping motion achieves its highest value. Interestingly, the optimal angles differ between the two regimes. Further, the periodic motion is observed only in a specific range of illumination angles. The range for $D<D^c$ is slightly larger than that for $D^c<D$. Finally, note that we also ran additional simulations at angle $\phi=30^\circ$, and found that $D^c=0.56$, indicating that critical penetration depth differs at different illumination angles. 

Finally, we comment on the role of $\lambda$.  In all these calculations, we have taken $\lambda >0$ consistent with the alignment of directors along the length of the strip.  However, if the alignment is perpendicular to the surface of the strip, then $\lambda <0$. In this case, the motion would be reserved (see Fig. S1 and Movie 3 in Supplement).\\ 

\noindent \textbf{Data accessibility:} All algorithms and data necessary for the analysis are included in the paper and supplementary materials.\\

\noindent \textbf{Competing interests:} The authors declare that they have no competing interests.\\

\noindent \textbf{Acknowledgment:} We gratefully acknowledge the support of the US Office of Naval Research through Multi-investigator University Research Initiative Grant ONR N00014-18-1-2624.\\

\bibliographystyle{unsrt}
\bibliography{reference}

\newpage
\renewcommand\thefigure{S\arabic{figure}}
\setcounter{figure}{0}
\renewcommand\thepage{S\arabic{page}}
\setcounter{page}{1}

\begin{center}
\text{Supplementary Material for:} \\
{\textbf{Optical penetration depth and periodic motion of a photomechanical strip}}\\
\vspace{0.5cm}
\text{\small Ameneh Maghsoodi}\\
\text{\small Department of Aerospace and Mechanical Engineering, University of Southern California}\\
\vspace{0.2cm}
\text{\small Kaushik Bhattacharya}\\
\text{\small Division of Engineering and Applied Science, California Institute of Technology}
\end{center}

\begin{figure}
\begin{center}
  \includegraphics[scale=0.85]{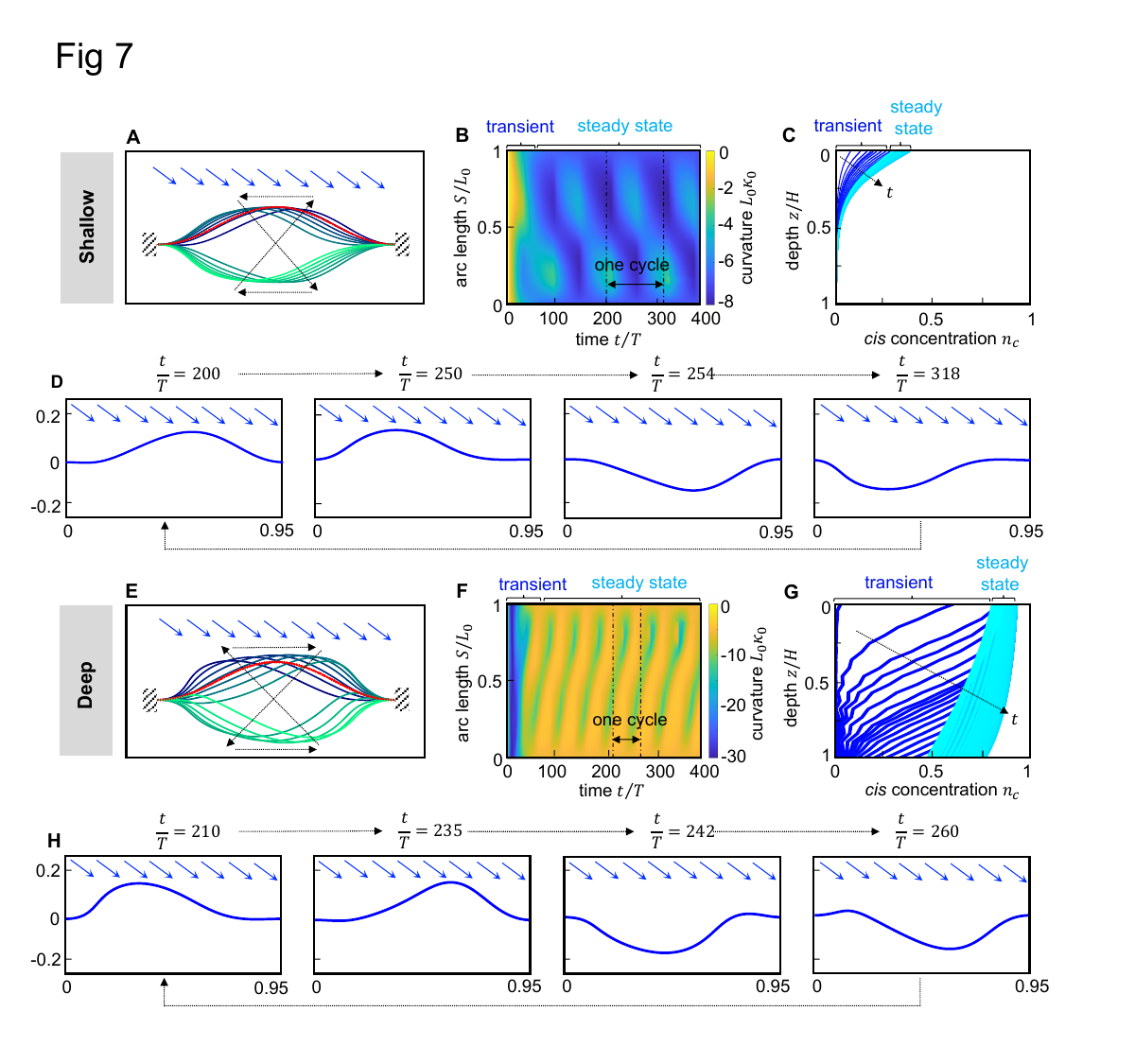}
  \caption{Dynamics of the photomechanical response of a doubly-clamped buckled strip with homeotropic molecular alignment. (A-D) correspond to shallow penetration ($I_0=630$W/m$^2$ and $D=0.2$), and (E-H) correspond to deep penetration ($I_0=31$kW/m$^2$ and $D=0.65$). (A) and (E) present the flapping motion per cycle; frequency $f=1.5$Hz for the shallow penetration, and frequency $f=2.3$Hz for the deep penetration. The red curves indicate the initial conformation of the strip when the light is off. (B) and (F) present the time evolution of non-dimensional spontaneous curvature $L_0 \kappa_0$ along the strip arc length. (C) and (G)  present the time evolution of $cis$ concentration $n_c$ across the thickness. (D) and (H) present the snapshots of light-induced deformations of the strip during one cycle. In these simulations, all parameters are the same as those in Fig. \ref{fig:CC} except for $\lambda= -0.025$. }
  
\label{fig:S1}
\end{center}
\end{figure}

\vspace{0.8cm}
\noindent Figure \ref{fig:S1} investigates the role of molecular alignment on the periodic motion of the doubly-clamped buckled strip. The strip possesses molecular alignment perpendicular to the surface of the strip, known as homeotropic alignment; then $\lambda <0$. We consider the ratio of homeotropic $\lambda$ to homogeneous $\lambda$ about -0.5 ignoring the volume loss during the illumination \cite{van2007glassy}. As shown in Figure \ref{fig:S1}, after an initial transient, the homeotropic strip undergoes a periodic flapping motion in the opposite direction of the homogeneous strip in Fig. \ref{fig:CC}. Also, the length of the strip increases at the deep penetration limit because of the axial extension for the homeotropic alignment.

\end{document}